\documentclass[twoside]{dis04}
\def\runauthor{\rm V.Radescu {\em et al.}}
\def\shorttitle{NuTeV Results}

\begin{document}

\title{Differential Cross Section Results from  NuTeV}

\author{Voica A. Radescu}
%\runauthor
\address{\rm for the NuTeV Collaboration\\
\it University of Pittsburgh,\\
Pittsburgh, PA 15260, USA\\
E-mail: voica@farfalle.phyast.pitt.edu }

\maketitle

\abstracts{The NuTeV experiment has collected high statistics,
high energy samples of $\nu$ and $\overline{\nu}$
charged-current interactions using the sign-selected Fermilab
neutrino beam. NuTeV has extracted $\nu$
and $\overline{\nu}$ differential cross sections
for DIS single-muon production at moderate $x>0.015$ and
average $Q^2\sim 15$~GeV$^2$. Differential cross sections
and structure function results are presented.
The NuTeV measurement has improved systematic
precision and includes data over an expanded
kinematic range up to high inelasticity.}

\section{Introduction}

Neutrino deep inelastic scattering (DIS) is a versatile probe of
nucleon structure.
The general form of the differential cross sections for the
neutrino--nucleon charged-current process depends upon three Lorentz-invariant
structure functions that parametrize the structure of the nucleon: $2xF_1,~
F_2,~ xF_3$:
\begin{equation}
  \frac{d^2\sigma^{\nu(\overline{\nu})}}{dx~ dy} =
  \frac{G_F^2 M
    E_\nu}{\pi}
\left [(1-y)F_2^{\nu(\overline\nu)}
+\frac{y^2}{2}2xF_1^{\nu(\overline\nu)}
\pm y\left ( 1-\frac{y}{2}\right )xF_3^{\nu(\overline\nu)}
\right ]
\label{eq1}
\end{equation}
where $G_F$ is Fermi weak coupling constant, $M$ is the nucleon mass,
$E_\nu$ is the incident neutrino energy, and $y$ the
inelasticity. Structure function $xF_3$ is unique to neutrino
interactions because neutrinos interact only weakly.
Structure functions depend on $x$, the Bjorken scaling variable
and $Q^2$, the square of the four-momentum transfer to the nucleon.
In neutrino scattering the Lorentz-invariant kinematic variables $x$, $y$
 and $Q^2$ are constructed from three measured quantities: the
 momentum of the outgoing muon, $p_\mu$, the angle of the outgoing muon
 with respect to the beam direction, $\theta_\mu$, and the energy of
 the outgoing hadrons, $E_{had}$.

NuTeV is a fixed target deep inelastic neutrino-scattering experiment 
which collected data during 1996-97 (Fermilab) using an
iron-scintillator neutrino detector \cite{nutcal}.
There are two important and improved features which make NuTeV the
most precise experiment to date. First unique feature is the
use of Sign Selected Quadrupole Train (SSQT) beam to produce a high
purity 
neutrino or, alternatively, anti-neutrino beam. 
The other special feature is 
the use of a continuous calibration beam
running concurrently with data-taking, 
which enables the 
NuTeV experiment to considerably improve its knowledge of the energy scale
and detector response functions. Muon energy scale was measured to
0.7\% and hadron energy scale to 0.43\%.

\section{Cross Section Measurements and Comparisons}

The differential cross section per nucleon on iron as function of $x$, $y$ 
and $E_\nu$ can be expressed in terms of the relative flux as function of 
energyand differential number of chargedcurrent events:
\begin{equation}
  \frac{d^2\sigma^{\nu(\overline{\nu})}(E_\nu)}{dx~ dy} =
 \frac{1}{\Phi(E_\nu)}\frac{d^2 N^{\nu(\overline{\nu})}(E_\nu)}{dx~ dy}. 
\label{eq2}
\end{equation}

The data selection criteria for the main cross section sample requires
a good muon track for accurate momentum measurement, event containment
and minimum energies thresholds: $E_{had}>10$ GeV, $E_{\mu}>15$ GeV
and $E_\nu>30$ GeV.
 To minimize
the effects of the non-perturbative contributions kinematic cuts of
$Q^2 >1 \rm GeV^2$  and $x<0.70$ are required.

The SSQT allows NuTeV to expand the data sample
of toroid-analyzed events to include previously inaccessible data
of low energy muons which stop inside the detector. Those events are 
reconstructed using exclusively information from their energy
deposition in the target calorimeter.
This is a new sample and it is treated separately. For this sample we
require good containment and $E_\mu >4$ GeV.

Neutrino relative flux is determined from a nearly independent sample
at low hadronic energy ($E_{had}<20$ GeV) using the ``fixed $\nu_0$
method'': as $y=\frac{E_{had}}{E_\nu}\to0$ the integrated number of
events is proportional to the flux. The absolute flux is obtained
by normalizing the cross section to the world average $\nu - Fe$ cross
section value of $\sigma^\nu/E_\nu=0.677\pm 
0.014 \times 10^{-38} \rm cm^2/GeV$ \cite{pdg}.
A detector simulation is used  to account for acceptance and
resolution effects and employs as input a (LO)QCD inspired cross
section model which is iteratively fit to the data until convergence
occurs (within 3 loops).

\begin{center}
\begin{figure}[h]
\vspace*{7.8cm}
    \includegraphics{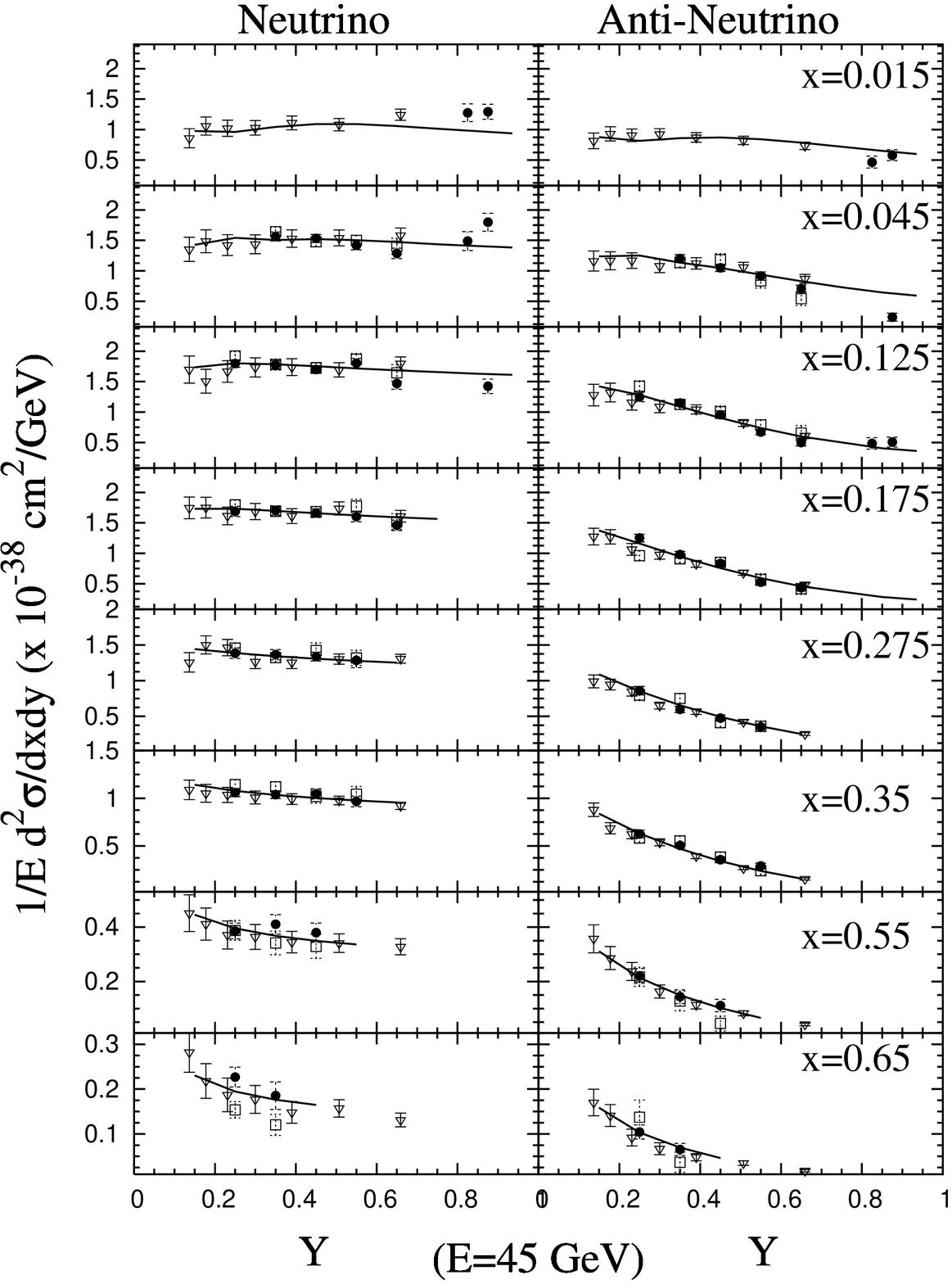}
    \includegraphics{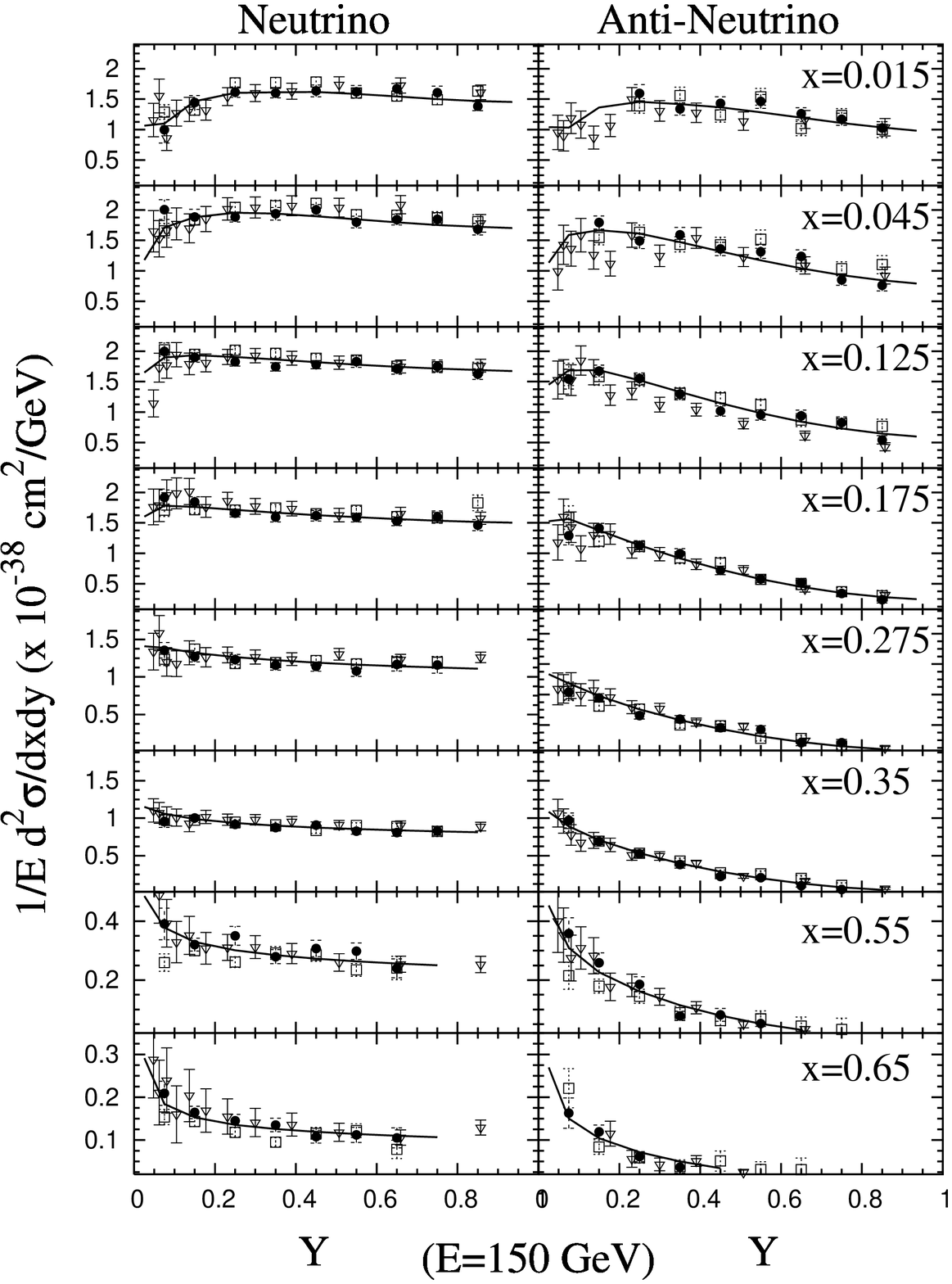}
    \vspace*{1.4cm}\caption[*]{ Plots display $\nu - Fe$ and
      $\overline{\nu}-Fe$ differential cross sections as function of $y$ at
      $E_\nu=45 GeV$ (left) and $E_\nu=150 GeV$ (right) over various $x$
      bins. NuTeV data over the extended $y$ region is shown with
      filled circles, CCFR set is marked with open squares and
      CDHSW data is  displayed with open triangles. The solid curve is the
      parametrization fit to NuTeV data. The main systematics are included 
      for all 
      NuTeV data points apart from the new high-$y$ sample.}
\label{xsec}
  \end{figure}
\end{center}

Figure~\ref{xsec} shows the extracted differential cross sections for
$\nu - Fe$ and $\overline\nu -Fe$ as function of $y$ for a representative
sample of $x$ bins at $E_\nu=45$ GeV and $E_\nu=150$ GeV. The NuTeV data
over an extended $y$ region 
is compared to other neutrino measurements: CCFR \cite{ukthesis}  and
CDHSW \cite{cdhs}. The solid curve is the parametrization fit to the
NuTeV data used for calculation of acceptance and smearing
corrections \cite{bg}. 
The main systematics due to energies scales and flux determination
have been included for NuTeV major data sample, while they still need to be
evaluated for the new high-$y$ sample.

For low and moderate $x$ there is good agreement among the data sets
over whole energy and $y$ range in both neutrino modes. At high $x$
NuTeV is systematically above the CCFR measurement over the entire
energy range. This difference increases with $x$ reaching up to 20\% at
$x=0.65$. NuTeV result is similar in level, but different in shape with
CDHSW, a data set with large uncertainties. 
 The main difference between NuTeV and
CCFR experiments, which are very similar in design and analysis
method, is that NuTeV beam was sign selected; it had
separate running of neutrinos and, respectively, anti-neutrinos. Having
this advantage, NuTeV was always set to focus the muon from the primary
vertex, whereas CCFR had simultaneous neutrino and anti-neutrino
runs, so that the toroid polarity was reversed periodically to focus
either $\mu^+$ or $\mu^-$.

\section{Structure Function Measurements and
  Comparisons}
\vspace*{-0.15cm}
Structure functions can be extracted from fits to linear combination
of $\nu$ and $\bar\nu$ differential cross sections, which have been
corrected for 5.67\%  excess of neutrons over protons in the NuTeV iron
target and for QED radiative effects \cite{bardin}. 
In this analysis we have not included the cross section points from
the high-$y$ sample.
The sum of the neutrino and anti-neutrino differential cross sections
 for charged current interactions on an iso-scalar target is related to
the structure function $F_2(x,Q^2)$ by:
\begin{equation}
\frac{d^2\sigma^{\nu}}{dxdy}\!+\! 
      \frac{d^2\sigma^{\overline{\nu}}}{dxdy}\!=\!
\frac{2MG_F^2E_\nu}{\pi}\!\left[\!\left(\!1\!-\!y\!-\!\frac{Mxy}{2E}\!+\!
\frac{1\!+\!(\frac{2Mx}{Q})^2}{1\!+\! R_L}\frac{y^2}{2}\!\right)\!F_2\!+\!
 y(1\!-\!\frac{y}{2})\Delta xF_3\!\right]\label{f2}
\end{equation}
where $R_L(x,Q^2)=\frac{\sigma_L}{\sigma_T}$ is the ratio of the cross
section for scattering from longitudinally to transversely polarized
-bosons, and $\Delta xF_3=xF_3^\nu-xF_3^{\overline\nu}\sim4x(s-c)$ is
sensitive to heavy flavors.
The structure function $F_2$ is determined by performing a one-parameter
fit which requires input models for $R_L(x,Q^2)$  and $\Delta
xF_3(x,Q^2)$. For  $R_L(x,Q^2)$ we use a fit to
the world's data \cite{rworld}, and for $\Delta xF_3(x,Q^2)$  we use
a NLO QCD model of Thorne-Roberts VFS \cite{trvfs}. 

Similarly, the difference between neutrino and 
anti-neutrino differential cross sections is related to the structure
function $xF_3(x,Q^2)$ as follows:
\begin{equation}
\frac{d^2\sigma^{\nu}}{dx~ dy} - 
      \frac{d^2\sigma^{\overline{\nu}}}{dx~ dy}=
\frac{2MG_F^2E_\nu}{\pi}\left(y-\frac{y^2}{2}\right)xF_3(x,Q^2)
\label{f3}
\end{equation}
Because $F_2^\nu(x,Q^2)\approx F_2^{\overline\nu}(x,Q^2)$, no model inputs are
needed to extract $xF_3$ from one-parameter fit.

\begin{center}
\begin{figure}[h]
  \vspace*{6.5cm}
  \includegraphics{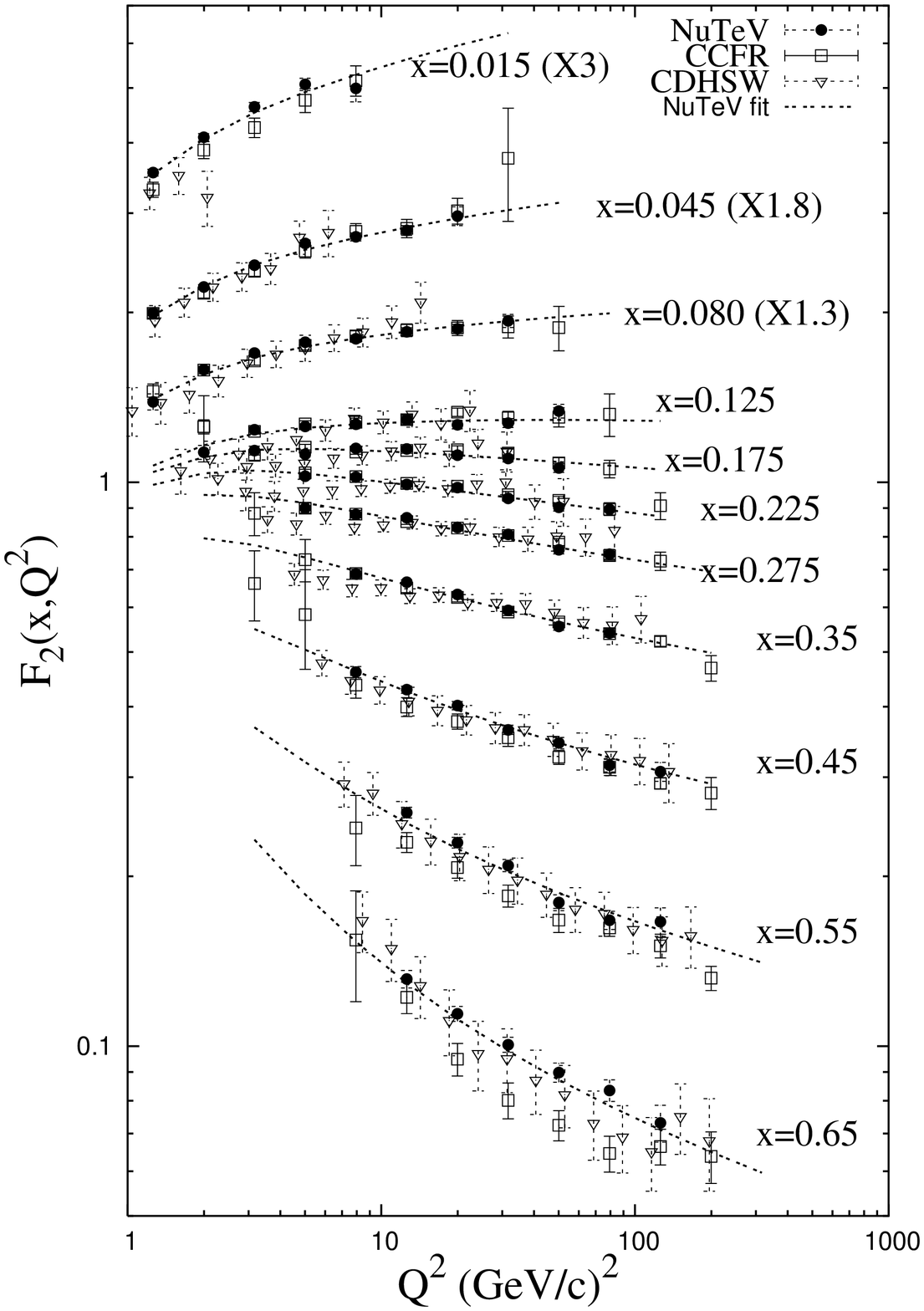}
  \includegraphics{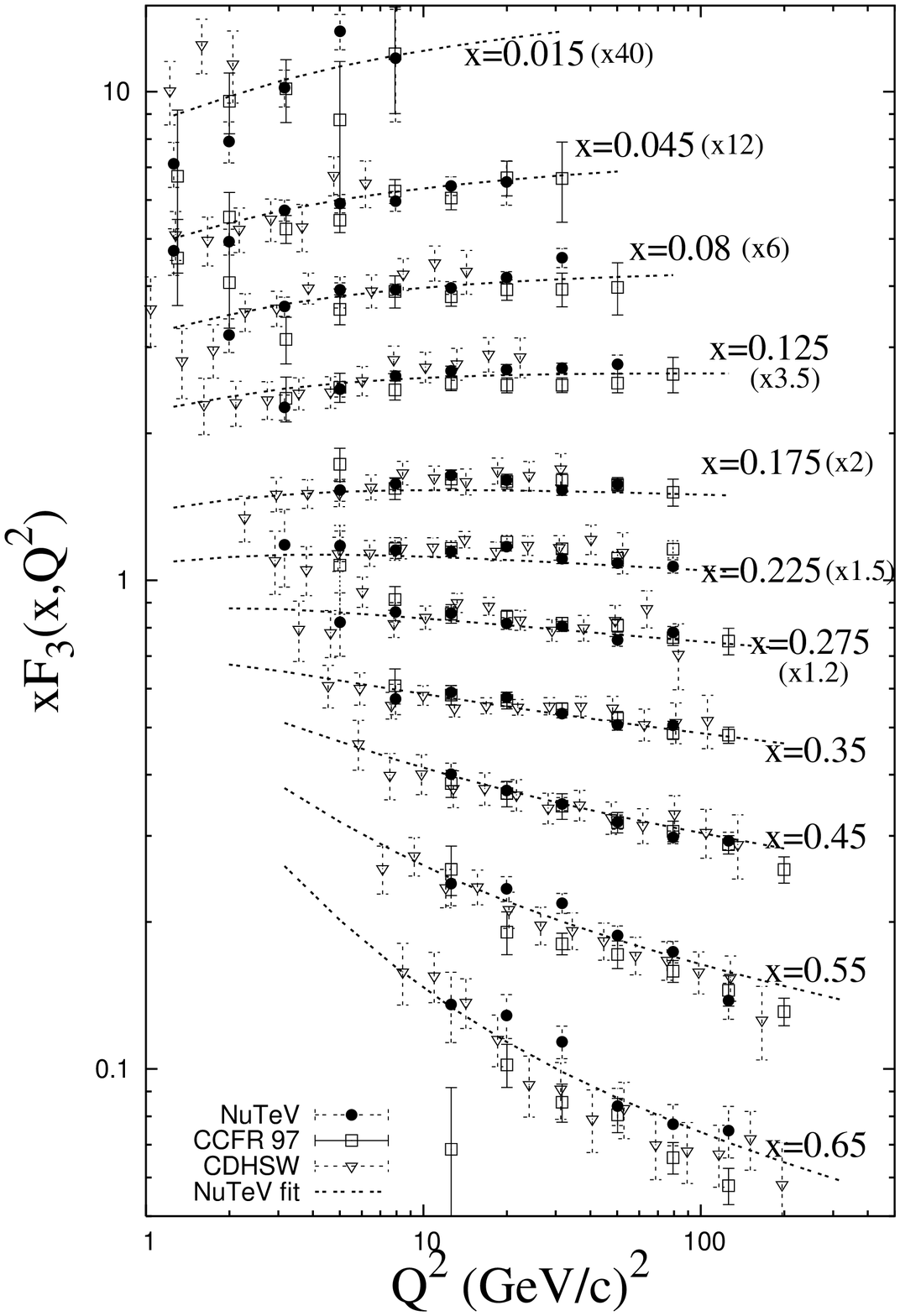}
   \vspace*{1.55cm}\caption[*]{Plots show NuTeV measurements of $F_2(x,Q^2)$ 
     (left) and $xF_3(x,Q^2)$ (right), filled circles, compared to CCFR,
     open squares, and CDHSW, triangles as function of $Q^2$ over all
     $x$ bins. Systematic errors are included. The curve is NuTeV model}
\label{sf}
\end{figure}
\end{center}

\begin{center}
\begin{figure}[h]
  \vspace*{8.75cm}
  \includegraphics{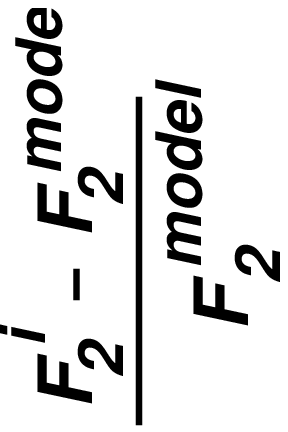}
  \includegraphics{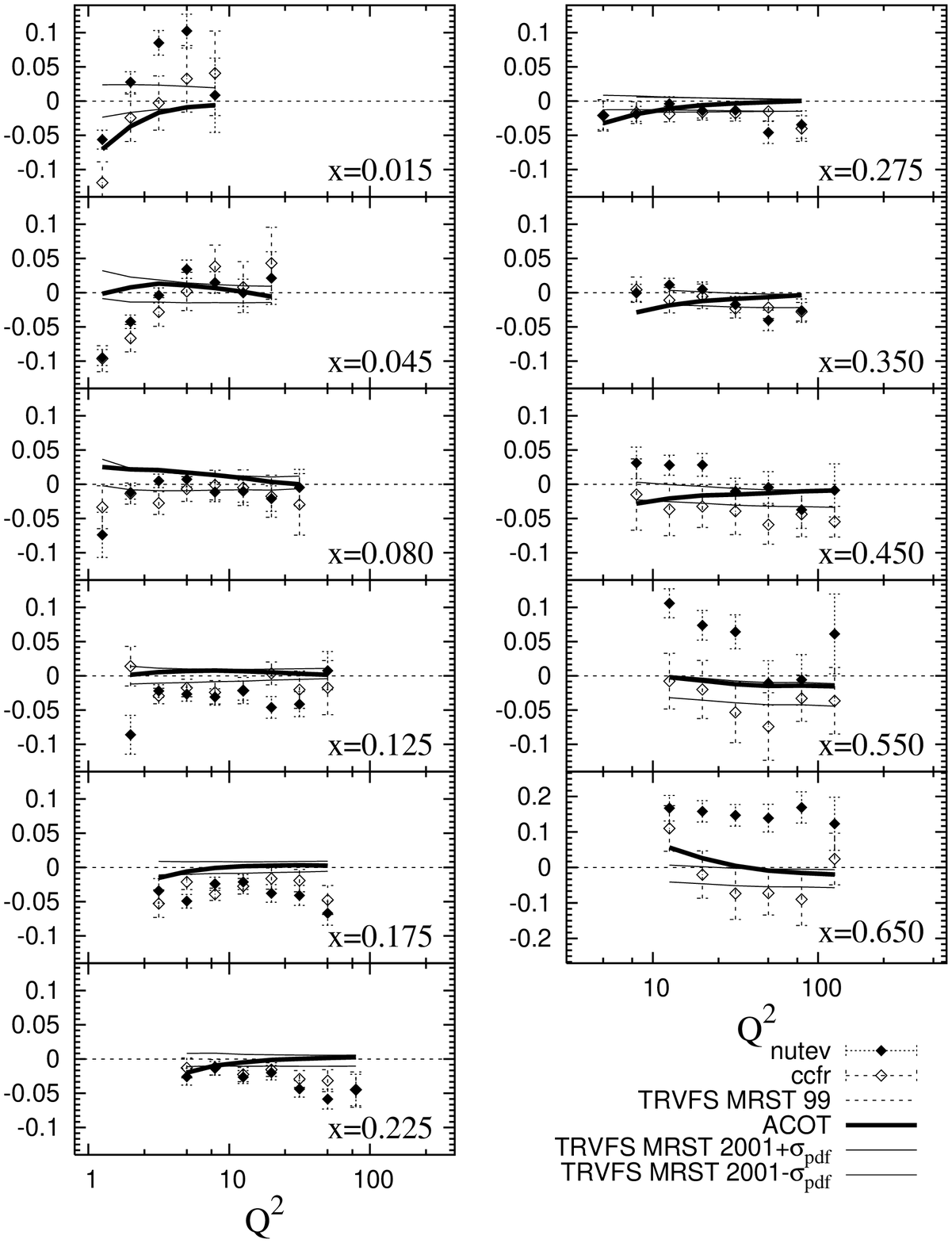}
  \vspace*{1.5cm}\caption[*]{Plots show ratios of the NuTeV (filled
    diamonds) and CCFR (open diamonds) $F_2(x,Q^2)$ measurements
    to TR-VFS(MRST-99 pdf) model (dashed line) as
    function of $Q^2$ over $x$ bins. The ratio of the
    ACOT (CTEQ4HQ pdf) (bold curve) and
    TR-VFS (MRST 2001 E pdf) with $\pm 1 \sigma$ 
    uncertainties in pdf's (thin curves) to the TR-VFS(MRST-99 pdf) model
    are also shown.}
\label{theo}
\end{figure}
\end{center}

Figure \ref{sf} shows comparisons of NuTeV measurements of
$F_2(x,Q^2)$ and $xF_3(x,Q^2)$ with those from CCFR and CDHSW. The
curve on the figure is the fit to NuTeV from our model. The major
systematics due to the $R_L$, $\Delta xF_3$ models, energies scales
and flux are included. The differences seen at the cross section level
are also reflected in the structure functions.
At $x=0.015$, the lowest $x$ bin, NuTeV is systematically above
CCFR by $\approx 3\%$. At the intermediate $x$, $0.015<x<0.5$, all
data sets are in good agreement. In the high $x$ region, $x>0.5$,
NuTeV is consistently above CCFR data up to $\approx 20\%$ at
$x=0.65$, and agrees in level, but not in shape with less precise
CDHSW data set. 
   \vspace*{-1.cm}
Figure \ref{theo} shows the ratios of the NuTeV and CCFR $F_2$
measurements to various NLO QCD models which include an improved
treatment of massive charm production: 
TR-VFS models \cite{trvfs} with MRST-99, MRST-2001 E parton
distribution functions sets and ACOT-FFS model
with CTEQ4HQ \cite{acot}. 
NuTeV data is significantly above the theory curves at
high-$x$ reaching $\approx 15\%$ difference at $x=0.65$, while the
CCFR result is slightly below the curves. 
\vspace*{-1.5cm}In order to compare neutrino measurements to the 
theory models, one must to correct the theory for target-mass effects \cite{tm}
and for nuclear effects which are important at low and high $x$.
 The nuclear corrections are measured in
charged lepton experiments from nuclear targets and the standard way
is to use the same correction for neutrino
scattering.
 We use a parametrization independent of $Q^2$ fit to data \cite{pdg} which is
 dominated at $x>0.4$ by SLAC,  a data set with lower
$Q^2$ than NuTeV.
 The size of the correction ranges from $\approx 10\%$ at
$x=0.015$, is small at intermediate $x$, and increases from $\approx 7\%$
at $x=0.45$ to $\approx 15\%$ at $x=0.65$.

\begin{figure}[h]
\vspace*{4.85cm}
    \includegraphics{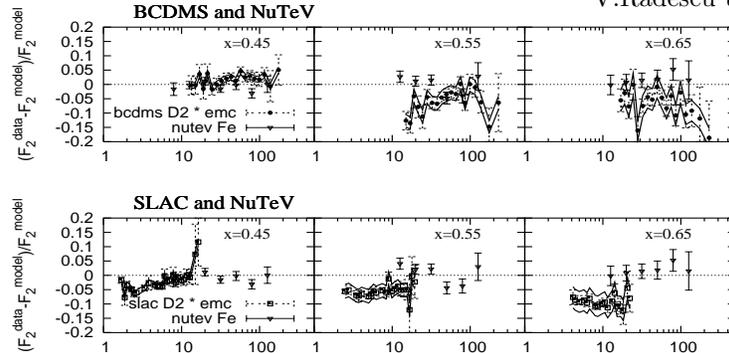}
\vspace*{-1.cm}\caption[*]{Ratio plots of NuTeV (triangles) and BCDMS
  $D_2$ (squares) $F_2$ measurements to NuTeV model are displayed as
  function of $Q^2$ for high-$x$ bins. The ratio plots of NuTeV
  (triangles) and SLAC $D_2$ (circles) $F_2$ measurements to NuTeV
  model are also shown.The error band on the charged
  lepton data is the uncertainty of the nuclear correction.}
\label{ch}
  \end{figure}

Currently there is a new analysis underway  with JLAB data using the Nacthmann 
variable $\xi=2x/(1+\sqrt{1+4M^2x^2/Q^2})$, which favors
slightly smaller nuclear corrections \cite{arrington}.
 We also compare our measurements in the high-$x$ region
to SLAC and BCDMS deuterium data sets \cite{slac},\cite{bcdms}. 
In order to compare $F_2^{\nu N}$ from $\nu-Fe$
DIS to the $F_2^{lD}$ from $e(\mu)-D_2$ scattering it is
necessary to apply two corrections: for the
quark charges seen by the electromagnetic interaction versus the weak
interaction and for the difference in light versus
heavy target effects.
Figure \ref{ch} shows the ratios of the NuTeV model to each data set
at high $x$ region. NuTeV measurements differ from
BCDMS by $\approx 5\%$  and from SLAC by $\approx 10\%$ at $x=0.65$, 
which could
indicate that neutrino scattering favors smaller nuclear effects at high $x$.

\vspace*{-1.0cm}
\section{Conclusions}
\vspace*{-0.25cm}
We have presented the most precise measurements of neutrino and
anti-neutrino differential cross sections to date \footnote{The figures
  in this proceedings have been recently updated and they contain minor
  changes to the results presented at the conference.}. The
sign-selected beam allows NuTeV to include a previously inaccessible high-$y$
data sample in the cross section.
 The measurements are in good agreement with previous neutrino
results over the intermediate $x$ range, but NuTeV results are
systematically higher at high-$x$ over the entire energy and $y$ range. 
The NuTeV measurements of $F_2$
are also compared to various NLO QCD models and in the high $x$ region
the results are systematically above theory curves, from
$5$ to $15\%$. Assumptions for the nuclear corrections have been made
 when compared to the theory models and charged lepton data.
The preliminary result is available on NuTeV web-page \cite{web}. 

\end{document}